\documentclass[a4paper,jou,natbib,floatsintext]{apa6}

\usepackage[english]{babel}
\usepackage[utf8x]{inputenc}
\usepackage{amsmath,amssymb}
\usepackage{graphicx}

\usepackage{url}
\usepackage{breakurl}
\usepackage[breaklinks]{hyperref}

\newcommand*{\doi}[1]{\href{http://dx.doi.org/#1}{#1}}

\title{Bayes Factors can only Quantify Evidence w.r.t.\ Sets of Parameters,\\not w.r.t.\ (Prior) Distributions on the Parameter}
\shorttitle{Bayesian Hypotheses}
\author{Patrick Schwaferts, Thomas Augustin}
\authornote{Author information:\\[0.5em]
Patrick Schwaferts: patrick.schwaferts@stat.uni-muenchen.de\\
Thomas Augustin: thomas.augustin@stat.uni-muenchen.de\\[0.5em]
Methodological Foundations of Statistics and its Applications\\
Department of Statistics\\
Ludwig-Maximilians-Universität Munich\\
Ludwigsstraße 33, 80539 Munich, Germany\\}
\affiliation{Ludwig-Maximilians-Universität Munich, Germany}

\abstract{Bayes factors are characterized by both the powerful mathematical framework of Bayesian statistics and the useful interpretation as evidence quantification. Former requires a parameter distribution that changes by seeing the data, latter requires two fixed hypotheses w.r.t.\ which the evidence quantification refers to. Naturally, these fixed hypotheses must not change by seeing the data, only their credibility should! Yet, it is exactly such a change of the hypotheses themselves (not only their credibility) that occurs by seeing the data, if their content is represented by parameter distributions (a recent trend in the context of Bayes factors for about one decade), rendering a correct interpretation of the Bayes factor rather useless. Instead, this paper argues that the inferential foundation of Bayes factors can only be maintained, if hypotheses are sets of parameters, not parameter distributions.
In addition, particular attention has been paid to providing an explicit terminology of the big picture of statistical inference in the context of Bayes factors as well as to the distinction between knowledge (formalized by the prior distribution and being allowed to change) and theoretical positions (formalized as hypotheses and required to stay fixed) of the phenomenon of interest.}

\begin{document}
\maketitle

\section{Introduction}

Statistical hypotheses have always been sets of parameters in classic frequentist hypothesis tests. However, in the context of Bayes factors -- a prominent Bayesian method for hypothesis comparisons  \citep[][]{Jeffreys1961,Kass1995,Goenen2005,Rouder2009} -- it is argued that also the prior distribution on the parameter might be employed to represent the hypotheses (or ``models'') that should be contrasted against each other. 
This view was promoted primarily by \citet{Vanpaemel2010} in an attempt to turn one of the fundamental issues of Bayes factors, namely its prior sensitivity \citep[see e.g.][]{Kass1995,Sinharay2002, Liu2008,Kruschke2015}, from a limitation to a feature. By now, this view can be found within many other publications, sometimes rather explicitly, sometimes only implicitly (see e.g.\  \citet[][p.~364f]{Dienes2019}, \citet[][p.~228]{Etz2018},  \citet[][p.~5]{Heck2020}, \citet[][p.~16]{Morey2016}, \citet[][]{Rouder2016}, \citet[][p.~776, 780]{Tendeiro2019}, \citet[][]{Vanpaemel2012}). Even the authors of this paper were previously influenced by this view \citep[][]{Ebner2019}.
However, as will be outlined within this paper, the inferential foundation of Bayes factors is severely impaired when representing statistical hypotheses via parameter distributions. Instead, statistical hypotheses need to be sets of parameters only, even in Bayesian statistics.

In order to elaborate these considerations, an explicit terminology is outlined first, building on the framework by \citet{Kass2011}. Subsequently, updating consistency of Bayes factors is given a detailed account to determine conditions that lead to inconsistencies. Finally, the representation of hypotheses by sets of parameters and by prior distributions is assessed, respectively, showing that former does not suffer foundational issues, only latter does.

\section{Big Picture}

\subsection{Formalization and Interpretation}

\begin{figure}[ht]
	\centering
  \includegraphics[width=0.9\linewidth]{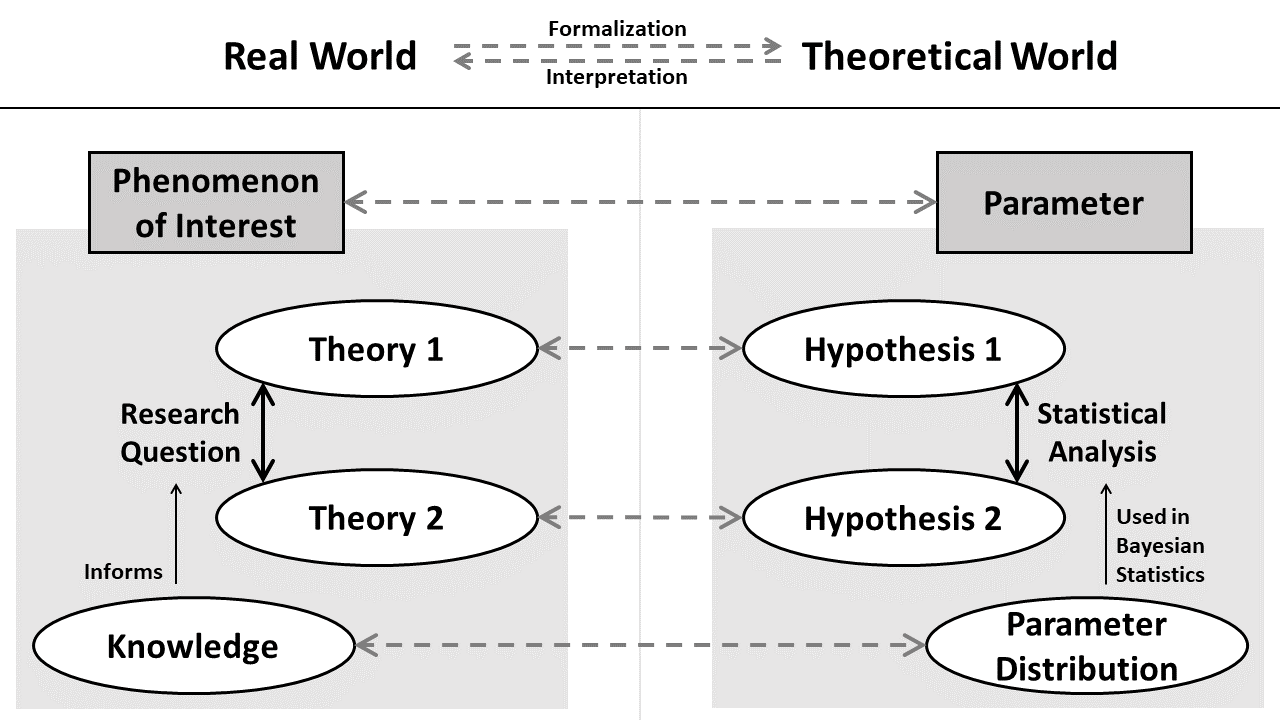}
	\caption{Big Picture of Statistical Inference in the Context of Bayes Factors. While the general view about the real world and the theoretical world is elaborated on by \citet[]{Kass2011}, the big picture of statistical inference in the context of Bayes factors is elaborated on in detail within this paper.}
	\label{fig:formalization}
\end{figure}

A comprehensive view of statistical inference distinguishes between the \textit{real world} and a \textit{theoretical world} \citep[][]{Kass2011}, where latter contains mathematical formalizations of the relevant characteristics in the real world. Interpreting the components of the theoretical world leads to their counterparts in the real world. In that sense, both worlds can be connected by \textit{formalization} and \textit{interpretation} (see Figure~\ref{fig:formalization}). Based on this general view by \citet[]{Kass2011}, the pig picture of statistical inference in the context of the Bayes factors shall be derived in detail in the following.

Typically, a researcher designs a scientific investigation to assess a phenomenon of interest. This scientific investigation leads to data that are described by a parametric sampling distribution, and the \textit{parameter} (which lives in the theoretical world) should correspond to \textit{phenomenon of interest} in the real world. If the correct interpretation of the parameter does not match with the phenomenon a researcher is interested in, then the design of the scientific investigation might be reconsidered. For the reminder of this paper, a proper correspondence between the parameter and the phenomenon of interest is assumed.

Within this fundamental framework, the big picture of statistical inference in the context of Bayes factors shall be established (as eventually depicted in Figure~\ref{fig:formalization}). This, however, is not easy, as relevant terms, as e.g.\ ``theory'', ``model``, or ``hypothesis'', have a multitude of different meanings and usages. In that, it is mandatory to explicitly define the employed terms such that their usage can be universally agreed on. Therefore, this elaboration should start at the very beginning, namely with two undeniable mathematical properties of Bayes factors.

\subsection{Common Ground: The Bayes Factor}

The Bayes factor (formulas below) is a quantity that is used within a \textit{statistical analysis} and lives in the theoretical world. Two mathematical properties of Bayes factors cannot be denied:
\begin{itemize}
\item It is a Bayesian quantity, such that it requires a distribution on the parameter.
\item It has a contrasting nature and contrasts two mathematical objects against each other (frequently referred to as hypotheses or models) \citep[cp.\ e.g.][Element~\#2 on p.~16]{Rouder2016}.
\end{itemize}
Put the other way: Without a parameter distribution or without contrasting two mathematical objects (hypotheses or models) there cannot be a Bayes factor.
Accordingly, whenever a Bayes factor is employed it is safe to assume the existence of a parameter distribution and the existence of a contrast and its two contrasted objects. Although these two simple facts about Bayes factors might seem trivial, it is important to state them explicitly, as it is exactly these two properties that serve as the basis to derive the big picture of statistical inference in the context of Bayes factors. While it is expected that there is a common agreement upon these two facts, there might be different views about other concepts employed in the context of Bayes factor (e.g.\ about the nature of hypotheses). By starting the elaboration with these two facts that can be agreed on, it is possible to assess the origin of disagreements about other concepts.

\subsection{Parameter Distribution and Knowledge}

\textit{Parameter distributions} (e.g.\ prior or posterior) live in the theoretical world and are typically interpreted as knowledge \citep[see e.g.][]{Jaynes2003} or uncertainty \citep[see e.g.][]{Kruschke2015} or degrees of belief \citep[see e.g.][]{Jeffreys1961} or information \citep[see e.g.][]{Berger1985} about the phenomenon of interest. Within this paper, the term knowledge shall be employed, as the exact label is not relevant for the elaborations below, only the fact that it is the interpretation of the parameter distribution. \textit{Accordingly, define the term knowledge (about a phenomenon of interest) within this paper as the interpretation of a Bayesian parameter distribution}.

\subsection{Hypotheses (or Models)}

The mathematical objects in the theoretical world that are contrasted by the Bayes factor shall be referred to as (statistical) \textit{hypotheses} (although the attribute ``statistical'' will e omitted as the term hypothesis is not employed in a non-statistical sense within this paper).
Other publications \citep[e.g.][]{Rouder2016, Rouder2018b} might state that the Bayes factor contrasts two \textit{models} against each other, yet the formula of the Bayes factor is exactly the same as in those publications that contrast hypotheses against each other \citep[cp.\ also][p.~11]{Morey2016}. Therefore, these models are the same mathematical objects as the hypotheses within this paper (namely those mathematical objects that are contrasted against each other by the Bayes factor). Other authors \citep[e.g.][]{Kruschke2018b} use both terms (model and hypothesis) rather interchangeably \citep[cp.\ also][p.~775, esp.\ footnote~1]{Tendeiro2019}. In the remainder of this paper, the term ``model'' shall be avoided, as it appears to have a variety of different other usages as well. \textit{Accordingly, define the term hypothesis within this paper as one of the two mathematical objects that are contrasted against each other by the Bayes factor}.

To assess the nature of this mathematical object is the aim of this paper and it will be argued that it can only be a set of parameters and not a parameter (prior) distribution.

\subsection{Theoretical Positions (or Theories) and Research Question}

The Bayes factor contrasts two mathematical objects against each other in the theoretical world, and the same scheme applies to the real world after interpretation: There is a contrast between two \textit{theoretical positions} about the phenomenon of interest in the real world. \textit{In that sense, define the term theoretical position within this paper as the interpretation of a hypothesis}.

The respective \textit{research question} contrasts these two theoretical positions against each other. Please note that, in general, the nature of potential research questions about the phenomenon of interest is extremely versatile. However, only those research questions can be answered by Bayes factors, that contrast two theoretical positions against each other.
If a research question contrasts two theories against each other, which cannot be formalized as those mathematical objects that are contrasted by the Bayes factor, then the Bayes factor is not suitable to answer such a research question.

Typically, the term ``theory'' is employed instead of ``theoretical position'', and it is said that the Bayes factor compares two ``theories'' \citep[cp.\ also][who use both terms]{Rouder2018b}.
In this context, both terms (theory or theoretical position) denote the same, namely the interpretation of a hypothesis (i.e.\ the interpretation of the mathematical objects that are contrasted against each other by the Bayes factor).
However,  the term ``theory'' might be used in a multitude of different other ways as well, e.g.\ in a non-contrasting context or such that it cannot be formalized as a hypothesis in the context of Bayes factors.
To avoid confusion and to emphasized its contrasting nature, only the term ``theoretical position'' shall be employed within this paper.

\subsection{Summary Terminology}

So far, the concepts of the big picture of statistical inference in the context of Bayes factor have been outlined and it should be emphasized that the terms ``hypothesis'', ``theoretical position'', and ``knowledge'' are used within this paper to facilitate an understandable elaboration. In fact, it might have been possible to merely use the descriptions ``mathematical objects that are contrasted against each other by the Bayes factor'' (hypotheses), ``interpretation of these mathematical objects'' (theoretical positions), and ``interpretation of a parameter distribution'' (knowledge). Together with the two above mentioned undeniable properties about Bayes factors, namely the existence of a contrast (of two mathematical objects) and the existence of a Bayesian parameter distribution, the employed concepts should have been explicitly outlined. Other publications might employ a different terminology, such that it is necessary to check which concepts are actually referred to by each term in each publication.

\subsection{Statistical Inference with Bayes Factors}

Statistical inference is the procedure of deriving conclusions from observed data. Naturally, there is a variety of different inferential approaches, each using different principles to extract information from the observed data.
The elegance of Bayes factors might be attributed to the fact that they combine two different approaches to statistical inference in one single quantity: Bayesian learning and evidential quantification.
\begin{itemize}
\item Within the Bayesian approach to statistical inference, a parameter prior distribution gets updated to a parameter posterior distribution by including the information from the observed data via Bayes rule. Conclusions are then derived solely from the parameter distribution.
\item Within the evidential approach to statistical inference \citep[cp.\ e.g.][]{Berger1988, Royall1997,Royall2004,Blume2011}, the information within the data are used to quantify evidence w.r.t.\ two different fixed theoretical positions about a phenomenon of interest. Assume two theoretical positions $A$ and $B$ are of interest (and specified in the research question) and assume the evidence within the data is quantified to be $5$, then the evidential interpretation is: After observing the data the credibility of $A$ over $B$ is $5$-times higher than before the data were observed \citep[see e.g.][]{Morey2016}
\end{itemize}
On the one hand, while Bayesian statistics is able to answer also different research questions, by using Bayes factors the nature of potential research questions is limited to those that contrast theoretical positions, thus allowing a useful and intuitive interpretation in the context of evidential quantification. On the other hand, while the framework of evidential quantification might consider a variety of different contrasting statistical analysis, by using Bayes factors the statistical analysis is restricted to the Bayesian framework, providing a thorough and powerful mathematical foundation \citep[see e.g.][]{Jeffreys1961,Berger1985}.

\subsection{Knowledge vs.\ Theoretical Positions}

Accordingly, to consider the inferential foundation of Bayes factors comprehensively, both the knowledge about the phenomenon of interest (Bayesian inference) and the theoretical positions about the phenomenon of interest (evidential inference) need to be distinguished. These concepts are fundamentally different! While Bayesian learning allows (or even requires) the knowledge itself to be altered, theoretical positions in the context of evidential quantification stay fixed, only their credibility may change. 
Without such a clear distinction, the inferential foundation of Bayes factors might break apart:
\begin{itemize}
\item If the theoretical positions themselves (not only their credibility) are allowed to change by observing the data, then the useful and intuitive interpretation as evidence quantification is lost. Assume two theoretical positions $A$ and $B$ are of interest (and specified in the research question) but change by seeing the data to the theoretical positions $C$ and $D$, respectively, and assume the evidence within the data is quantified to be $5$, then the correct but useless interpretation is: The credibility of $C$ over $D$ after the data were observed is $5$-times higher than the credibility of $A$ over $B$ before the data were observed.
\item If the knowledge (and thus the parameter distribution) is forced to stay fixed although some data were observed, then Bayes rule is not applied, leading to updating inconsistencies (outlined in detail below).
\end{itemize}

Accordingly, a clear distinction between knowledge (formalized as parameter distributions and being allowed to change) and theoretical positions (formalized as statistical hypotheses which stay fixed) about a phenomenon of interest is mandatory.
In that, the framework depicted here (Figure~\ref{fig:formalization}) does account for the fundamental different nature of knowledge and theoretical positions about a phenomenon of interest.

Interestingly, on a side note, it might be stated that -- by its nature -- also the prior knowledge is able to inform the research question. Typically, prior knowledge is insufficient to answer the research question adequately, justifying the necessity to conduct a scientific investigation. However, the structure of how to answer the research question with the available knowledge is independent of whether data were observed or not: In a Bayesian context, it is a parameter distribution from which the answer to the research question is derived, and this way of deriving answers might work both for the prior and the posterior distribution.

\subsection{One-to-One Correspondence}

Ideally, the correspondence between the concepts of the real world with those in the theoretical world (gray dashed arrows in Figure \ref{fig:formalization}) should be one-to-one, mathematically described by a bijective mapping. Without such a bijective mapping, chosen formalizations might be arbitrary or the interpretation of the results might inform past the research question.
While the correspondence between the phenomenon of interest and the parameter depends on the quality of the experimental design, and the mapping between parameter distributions and knowledge is typically assumed to be bijective in the Bayesian setting (two different distributions represent two different bodies of knowledge), of interest for this elaboration is the relation between theoretical positions and hypotheses. Consider two cases:
\begin{itemize}
\item There is a bijective mapping between theoretical positions and hypotheses. Then, different hypotheses represent different theoretical positions.
\item A variety of different hypotheses formalize one single theoretical position. When forced to commit oneself to one of those hypotheses (as typically required by the statistical analysis), this choice might intuitively be called instantiation: The theoretical position is formally instantiated by one of the hypotheses. This terminology is frequently employed it the literature that (potentially implicitly) assume hypotheses to be represented by distributions (see e.g.\ \citet[p.~13]{Morey2016}, \citet[p.~2]{Rouder2018b}, \citet[p.~491]{Vanpaemel2010}, \citet[p.~1054]{Vanpaemel2012}), suggesting that such a non-bijective relation might be implicitly assumed. Other publications do also employ such a non-bijective relation without using the term instantiation \citep[e.g.][Box~3 on p.~369]{Dienes2019}.
\end{itemize}
In that, these two types of relations between theoretical positions and hypotheses might describe the ideal and the actual situation, respectively, and will be used below for elaborating issues inherent to representing hypotheses by parameter distributions.

\section{Updating Consistency}

\begin{figure}[ht]
	\centering
  \includegraphics[width=0.9\linewidth]{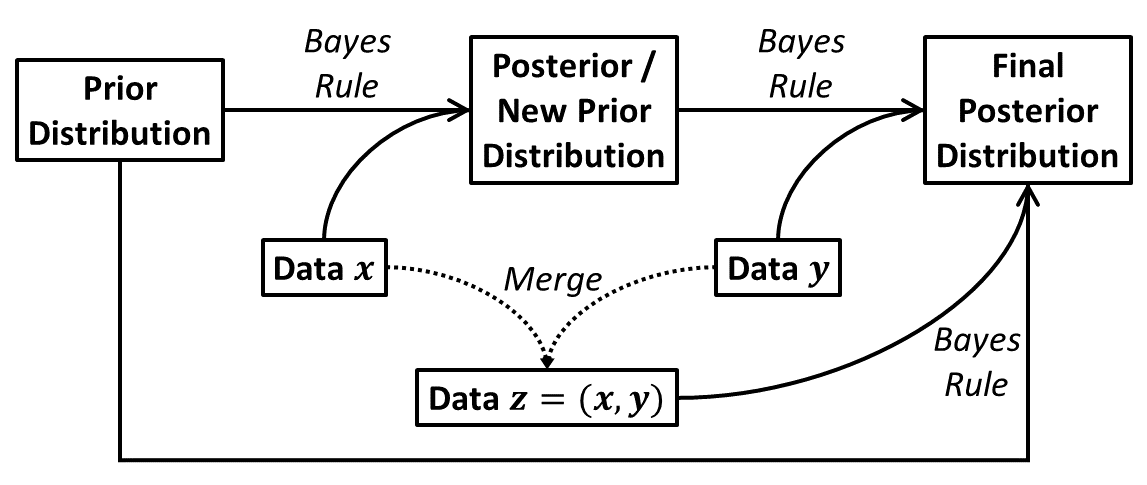}
	\caption{Consistent Bayesian Updating. Updating subsequently with two independent data sets $x$ and $y$ (top path) should yield the same final posterior distribution than merging both data sets first and then updating at once (bottom path).}
	\label{fig:updating}
\end{figure}

In Bayesian statistics, it is Bayes rule and not another principle, which states how a prior distribution gets updated to a posterior distribution (i.e.\ how information is extracted from the observed data; see Figure \ref{fig:updating}, top-left).
If the prior distribution reflects all available knowledge about the phenomenon of interest \textit{before} the investigation is conducted, then the corresponding posterior distribution reflects all available knowledge about the phenomenon of interest \textit{after} the investigation is conducted.
Disagreeing would imply that Bayes rule is not able to extract all information within the observed data that is relevant for the phenomenon of interest (and no Bayesian would do so).
As a consequence, the posterior distribution might be employed as a prior distribution for the Bayesian analysis of a data set obtained in a subsequent investigation with the same design (see Figure \ref{fig:updating}, top-right). Naturally, the final posterior distribution after subsequently updating twice should be identical to the posterior distribution obtained by merging both data sets first and then updating the initial prior distribution at once (see Figure \ref{fig:updating}, bottom). If so Bayesian updating is \textit{consistent} \citep[cp.][p.~190]{Ruger1998}, else it is \textit{inconsistent}.

\begin{figure}[ht]
	\centering
  \includegraphics[width=0.9\linewidth]{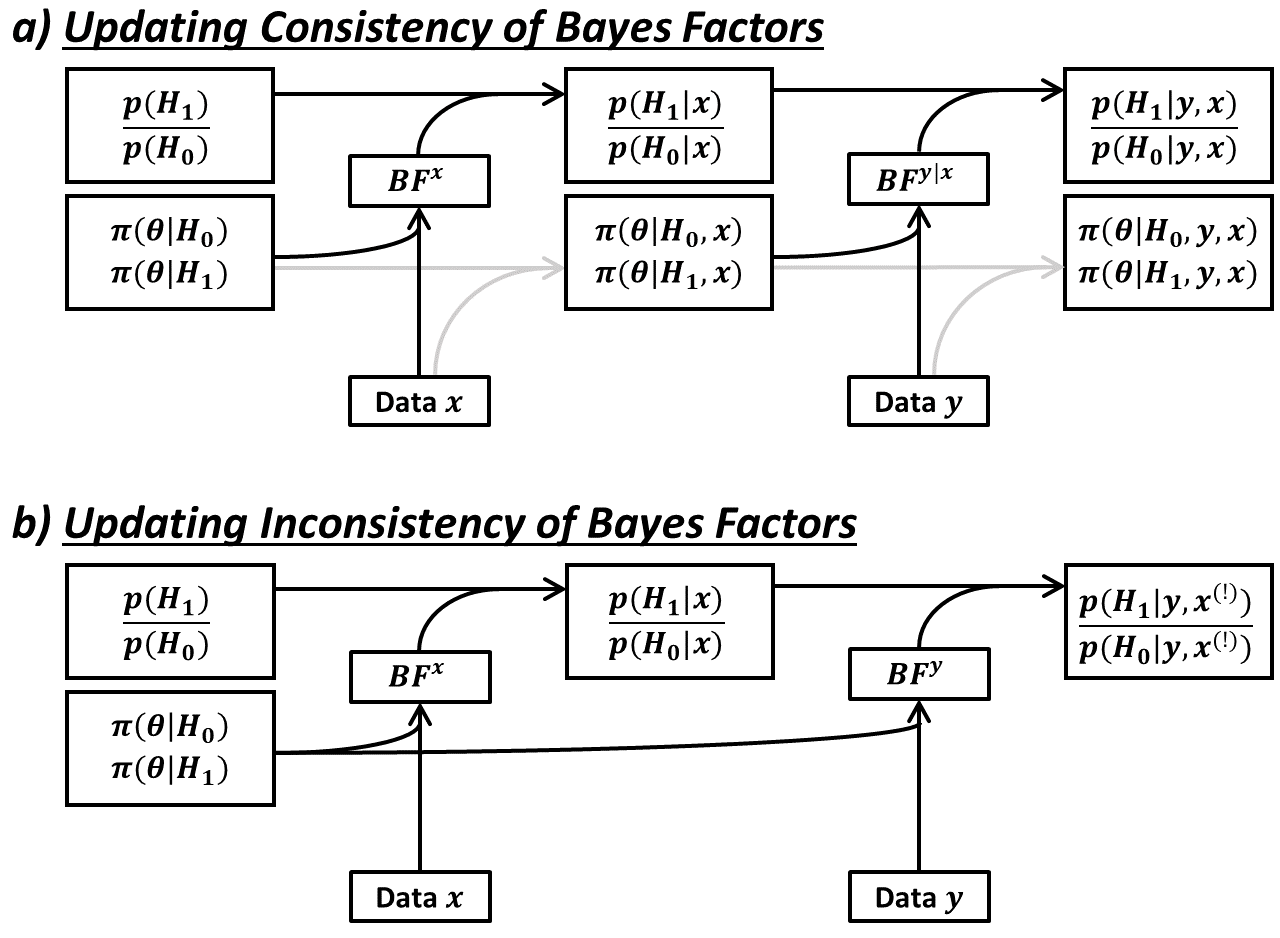}
	\caption{Consistent (a) and Inconsistent (b) Updating with Bayes Factors. Superscript $(!)$ indicates that not all relevant information was extracted from the data set $x$.}
	\label{fig:updating_BF}
\end{figure}

In general, updating with Bayes factors is consistent \citep[][]{Schwaferts2021} (see Figure~\ref{fig:updating_BF}a). Assume the observed data $x = (x_1,\dots,x_n)$ consists of independent and identically distributed observations $x_i$ ($i=1,\dots,n$) that follow the parametric sampling distribution with density $f(x_i|\theta)$, where $\theta \in \Theta$ is the parameter (representing the phenomenon of interest), such that the density of the complete data set $x$ is $f(x|\theta) = \prod_{i=1}^n f(x_i|\theta)$. The hypotheses $H_0$ and $H_1$ have prior probabilities $p(H_0)$ and $p(H_1) = 1 - p(H_0)$, and the corresponding densities of the hypothesis-based parameter distributions are $\pi(\theta|H_0)$ and $\pi(\theta|H_1)$, respectively. Then the density of the overall prior distribution is \citep[see e.g.][]{Rouder2018}
\begin{equation}\label{eq:prior}
\pi(\theta) = p(H_0) \, \pi(\theta|H_0) + p(H_1) \, \pi(\theta|H_1) \, .
\end{equation}
The Bayes factor
\begin{equation}\label{eq:BF}
B\!F^x = \frac{\int f(x|\theta) \, \pi(\theta|H_1) \: d\theta }{\int f(x|\theta) \, \pi(\theta|H_0) \: d\theta}
\end{equation}
is calculated using only the data $x$ and the hypothesis-based parameter densities $\pi(\theta|H_0)$ and $\pi(\theta|H_1)$, and allows to update the prior probabilities of the hypotheses to their posterior probabilities (see Figure \ref{fig:updating_BF}a, left):
\begin{equation}
\frac{p(H_1|x)}{p(H_0|x)} = B\!F^x \cdot \frac{p(H_1)}{p(H_0)} \, .
\end{equation}
In addition, revealed by simply applying Bayes rule consistently to the overall prior density $\pi(\theta)$ \citep[depicted in detail by][]{Schwaferts2021}, also the hypothesis-based parameter densities $\pi(\theta|H_0)$ and $\pi(\theta|H_1)$ get updated by the data~$x$ to their posterior densities $\pi(\theta|H_0, x)$ and $\pi(\theta|H_1, x)$ (gray arrow in Figure \ref{fig:updating_BF}a, left) \citep[cp. also][]{Kruschke2018b}. In general (i.e.\ for non-degenerate prior distributions), these posteriors are different than the priors.
These updated hypothesis-based posterior densities together with the posterior probabilities on the hypotheses describe the overall posterior distribution:
\begin{equation}
\pi(\theta|x) = p(H_0|x) \, \pi(\theta|H_0, x) + p(H_1|x) \, \pi(\theta|H_1, x) \, .
\end{equation}
If a new data set $y$ was observed (using the same experimental setup, i.e.\ following the same sampling distribution), this updated posterior distribution describes the starting point for a subsequent analysis with Bayes factors (Figure \ref{fig:updating_BF}a, right). Consequently, the corresponding Bayes factor
\begin{equation}
B\!F^{y|x} = \frac{\int f(y|\theta) \, \pi(\theta|H_1, x) \: d\theta }{\int f(y|\theta) \, \pi(\theta|H_0, x) \: d\theta}
\end{equation}
is also inherently influenced by the information within the previous data set $x$. In that, however, updating is consistent \citep[a complete proof is provided by][]{Schwaferts2021}.

Updating inconsistencies occur, if the second data set $y$ is analyzed with the initial hypothesis-based prior distributions (i.e.\ $\pi(\theta|H_0)$ and $\pi(\theta|H_1)$ as in equation (\ref{eq:BF})) although the first data set $x$ was already observed (Figure \ref{fig:updating_BF}b). This happens, if the initial hypothesis-based prior distributions do not get updated in the analysis of the first data set $x$, which is a violation of Bayes rule \citep[cp. also][who noticed this issue and solved it properly by merging all data sets first]{Rouder2011}. 
It is difficult to assess the prevalence of such updating inconsistencies in the current scientific literature, as the data is typically analyzed at once (not subsequently), and the calculation of the Bayes factor can be done without explicitly updating the hypothesis-based priors. Nevertheless, this update must not be neglected to be consistent (Figure \ref{fig:updating_BF}a) and the extent to which this fact is overlooked might be indicated by \citet{Tendeiro2019}: After a year of literature review about Bayes factors to understand them, the authors (and possibly their reviewers from the journal \textit{Psychological Methods} as well) were convinced (see footnote~2 on p.~776 therein) that the hypothesis-based priors do not get updated to their posteriors. Interestingly, at the same place, the authors refer to \citet[][]{Kruschke2018b}, who, in contrast, elaborated on the update of the hypothesis-based priors rather explicitly (see p.~157f and Fig.~4 and 6 therein). This is a vivid sign of the confusion a researcher faces in the literature about Bayes factors.

\section{Hypotheses as Sets of Parameters}

At first, assume that hypotheses are represented by two disjoint subsets $\Theta_0, \Theta_1 \subset \Theta$ of the parameter space $\Theta$:
\begin{equation}
H_0: \theta \in \Theta_0 \quad \text{vs.} \quad H_1: \theta \in \Theta_1 \, .
\end{equation}
These subsets need to be chosen to correspond to the theoretical positions that are contrasted within the research question. In addition to these theoretical positions, there is knowledge about the phenomenon of interest that is formalized by a parameter distribution with density $\pi(\theta)$. Without loss of generality, assume that this distribution has a positive density only for parameter values that are contained in one of the hypotheses.
The prior probabilities of the hypotheses are obtained from this parameter distribution by
\begin{equation}\label{eq:prior_probs}
p(H_0) = \int_{\Theta_0} \pi(\theta) \: d\theta \quad \text{and} \quad p(H_1) = \int_{\Theta_1} \pi(\theta) \: d\theta\, ,
\end{equation}
and, once the data $x$ were observed and the prior density $\pi(\theta)$ was updated to $\pi(\theta|x)$, the posterior probabilities of the hypotheses are
\begin{equation}
p(H_0|x) = \int_{\Theta_0} \pi(\theta|x) \: d\theta \quad \text{and} \quad p(H_1|x) = \int_{\Theta_1} \pi(\theta|x) \: d\theta\, .
\end{equation}
How the data change the probabilities of the hypotheses is described by the Bayes factor
\begin{equation}
B\!F^x = \frac{p(H_1|x)}{p(H_0|x)} \: / \: \frac{p(H_1)}{p(H_0)} \, .
\end{equation}
Accordingly, the probabilities of the hypotheses change but the hypotheses themselves, i.e.\ the sets $\Theta_0$ and $\Theta_1$, stay the same. In that, the Bayes factor can be interpreted appropriately as evidence quantification.
Further, the overall prior distribution ($\pi(\theta)$) was updated completely to the overall posterior distribution $(\pi(\theta|x)$), providing updating consistency. Consequently, hypotheses can safely be represented by sets of parameters in the context of Bayes factors.

Before continuing, the current situation shall be characterized further (which will be needed below).
Any given prior distribution with density $\pi(\theta)$ formalizes knowledge about the phenomenon of interest. One part of this knowledge relates to the one and another part to the other theoretical position (which are contrasted within the research question), formalized by the hypothesis-based prior densities $\pi(\theta|H_0)$ and $\pi(\theta|H_1)$. Mathematically, these densities are obtained from the initial density $\pi(\theta)$ by
\begin{align}
\pi(\theta|H_0) &= \frac{1}{p(H_0)} \cdot \pi(\theta)|_{\Theta_0} \label{eq:pi_h0} \\
\pi(\theta|H_1) &= \frac{1}{p(H_1)} \cdot \pi(\theta)|_{\Theta_1} \, , \label{eq:pi_h1}
\end{align}
where $\pi(\theta)|_{\Theta_0}$ and $\pi(\theta)|_{\Theta_1}$ are the densities $\pi(\theta)$ restricted to the sets $\Theta_0$ and $\Theta_1$, respectively.
Now, consider the set $\mathcal{X}$ of all potentially observable data sets of any size $n \in \mathbb{N}_0$, with $n=0$ referring to the empty data set (representing the prior situation). For a given prior distribution with density $\pi(\theta)$, denote the set of all potentially obtainable posterior densities as
\begin{equation}
\Pi := \left\lbrace \pi(\theta|x) \mid x \in \mathcal{X} \right\rbrace \, .
\end{equation}
This set contains all possible posterior distributions, i.e.\ represents all possible bodies of knowledge about the phenomenon of interest, that might be available after some (yet unknown) data $x$ were observed.
Analogously, all different possible bodies of knowledge about the theoretical positions, respectively, are formally contained within the sets
\begin{align}
\Pi_0 &:= \left\lbrace \pi(\theta|H_0, x) \mid x \in \mathcal{X} \right\rbrace \label{eq:Pi0} \\
\Pi_1 &:= \left\lbrace \pi(\theta|H_1, x) \mid x \in \mathcal{X} \right\rbrace \, .\label{eq:Pi1}
\end{align}
These sets contain only probability distributions with probability mass in the sets $\Theta_0$ and $\Theta_1$, respectively.
In summary, a hypothesis and all potentially obtainable bodies of knowledge about this hypothesis (in the context of given prior knowledge and a certain experimental setup) can be described by the sets $\Theta_0$ and $\Pi_0$ or $\Theta_1$ and $\Pi_1$, respectively.

\section{Hypotheses as Parameter Distributions}

Now, prior distributions with densities $\pi(\theta|H_0)$ and $\pi(\theta|H_1)$ shall represent the hypotheses $H_0$ and $H_1$, respectively.
Ideally, the mapping between theoretical positions and hypotheses should be bijective, updating should be consistent, and theories should not change by seeing the data $x$, only their credibility should.
In the following, two of each of these properties shall be assumed first and then evaluated w.r.t.\ the third.

\subsection{Case 1: Bijective Mapping and Updating Consistency}

Assume there is a bijective mapping between theoretical positions and hypotheses. 
As hypotheses are represented by prior distributions, not only by a set of parameters, two different parameter distributions represent two different hypotheses, i.e.\ two different theoretical positions.
Consistent Bayesian updating dictates to update the prior distributions to the posterior distributions, which are (for non-degenerate cases) different than the respective prior distributions.
As the parameter distributions change by observing the data, so do the hypotheses and theoretical positions.
It that, observing data changes the theories that should be contrasted with each other, not only their credibility. This does not match with the fundamental characteristics of statistical inference by evidence quantification, leading to issues with the interpretation of Bayes factors.

\subsection{Case 2: Bijective Mapping and Unchanged Theories}

Again, assume there is a bijective mapping between theoretical positions and hypotheses, and that hypotheses are represented by the prior distributions, such that two different parameter distributions represent two different hypotheses, i.e.\ two different theoretical positions.
If theories should not change by observing data, the posterior distribution needs to be the same as the prior distribution, which is outlined in Figure~\ref{fig:updating_BF}b and leads -- for non-degenerate prior distributions -- to updating inconsistency. In that, inference does not follow Bayes rule.

\subsection{Case 3: Updating Consistency and Unchanged Theories}

Now, also allow the mapping between theoretical positions and hypotheses to be non-bijective, in a sense that a single theoretical position might be formalized by a multitude of different hypotheses. If hypotheses are represented by prior distributions, then a set of different distributions corresponds to one theoretical position. How does this set look like, if updating should be consistent and a proper evidential interpretation of Bayes factors shall be kept?

To answer this question, assume that in the specifications of a statistical analysis each of both theoretical positions is instantiated by only one prior distribution with density $\pi(\theta|H_0)$ or $\pi(\theta|H_1)$, respectively, such that their supports do not overlap with each other (overlapping hypotheses will be discussed subsequently). Denote their supports (i.e.\ the sets of parameters in which the density has non-zero, positive mass) with $\Theta_0$ and $\Theta_1$, respectively.
If updating shall be consistent, then parameter distributions are allowed to change by seeing the data. Considering all potentially observable data sets $x \in \mathcal{X}$ (of any size $n \in \mathbb{N}_0$), the initial prior densities $\pi(\theta|H_0)$ and $\pi(\theta|H_1)$ might result in any posterior density within the sets $\Pi_0$ and $\Pi_1$ (equations (\ref{eq:Pi0}) and (\ref{eq:Pi1})), respectively.
To keep the proper evidential interpretation of Bayes factors, all these parameter densities within the sets $\Pi_0$ and $\Pi_1$ need to represent the same theoretical position, respectively.
In that, the hypotheses $H_0$ and $H_1$ are represented by the sets $\Pi_0$ and $\Pi_1$, which, however, contain all potentially observable parameter distributions with positive probability mass restricted to the parameter sets $\Theta_0$ and $\Theta_1$, respectively.
Two different parameter distributions with the same support (either $\Theta_0$ or $\Theta_1$) do not differentiate between two theoretical positions, only two different supports, i.e.\ sets of parameters, do. Accordingly, this situation is practically equivalent to representing hypotheses as sets of parameters.

\subsection{Overlapping Hypotheses}

Within these elaborations, it was assumed that the hypotheses are non-overlapping. Mathematically, they might also be overlapping. Consider the case, in which $\Theta_0$ and $\Theta_1$ are not (almost everywhere w.r.t.\ the prior density $\pi(\theta)$) disjoint.
Then there are parameter values $\theta$ that are contained within both $\Theta_0$ and $\Theta_1$, such that, if these parameter values are true, the posterior distribution will be shifted -- for an increasing sample size~$n$ -- within this overlapping part. As a consequence, even an infinitely large data set cannot decisively distinguish between both hypotheses (i.e.\ answer the research question), and the Bayes factor has a finite limit.
Formally, the behavior\footnote{This equation (\ref{eq:BF_inf}) has been formulated in a mathematically imprecise way in order to present the relevant points clearly. Actually, the condition $\theta^* \in \Theta_0 \setminus \Theta_1$ constitutes the case in which the true parameter $\theta^*$ is within the set $\Theta_0 \setminus \Theta_1$ such that for increasing $n$ the posterior distribution will be shifted into this set $\Theta_0 \setminus \Theta_1$. The other conditions have to be read analogously. There might be cases in which -- mathematically -- $\theta^*$ is within one of the parameter sets that define the conditions, but the posterior distribution will not be shifted completely within the respective set for $n \to \infty$. These cases, however, lie exactly at the borders between the set-valued hypotheses and are expected to occur almost never w.r.t.\ the parameter distribution.} of the Bayes factor is
\begin{equation} \label{eq:BF_inf}
B\!F
\begin{cases}
\to 0 \quad \quad \text{if} \: \: \theta^* \in \Theta_0 \setminus \Theta_1 \\
\to \infty \quad \: \: \:  \text{if} \: \: \theta^* \in \Theta_1 \setminus \Theta_0 \\
\to c(\theta^*) \: \: \, \text{if} \: \: \theta^* \in \Theta_0 \cap \Theta_1
\end{cases} \text{for} \: \: n \to \infty \, ,
\end{equation}
where $\theta^*$ is the true parameter and $c(\theta^*)$ is a fixed value that depends on $\theta^*$ \citep[cp.][p.~411]{Morey2011}.

In that, if -- for a given investigational setup -- the theoretical positions (that are of interest in the context of the research question) are reasonably formalized by overlapping hypotheses, it might happen that the scientific investigation cannot answer the research question. This might be a waste of time and money, and cannot be argued to yield a ``strong inference'' \citep[][]{Platt1964} or constitute a ``severe test'' \citep[][]{Popper2002}, claims frequently raised by promoters of the Bayes factor (see e.g.\ \citet[][p.~365]{Dienes2019}, \citet[][p.~228]{Etz2018}, \citet[][p.~130]{Schonbrodt2018}), which apparently require non-overlapping hypotheses. In this context it shall be noted that even \citet[e.g.\ p.~269]{Jeffreys1961} himself tried to avoid the possibility of obtaining a finite, non-zero Bayes factor limit at all costs: His derivation of the prior distributions for Bayes factors (now sometimes referred to as default Bayes factors \citep[see e.g.][]{Ly2016}) aimed at being able to state decisive evidence (which corresponds to Bayes factors tending to $\infty$ or $0$) even for a fixed number of observations (which is an even stronger requirement than within equation (\ref{eq:BF_inf})). In this regard, it seems advisable to rethink the investigational setup such that overlapping hypotheses can be avoided.

\section{Example}

\begin{figure}[ht]
	\centering
  \includegraphics[width=0.9\linewidth]{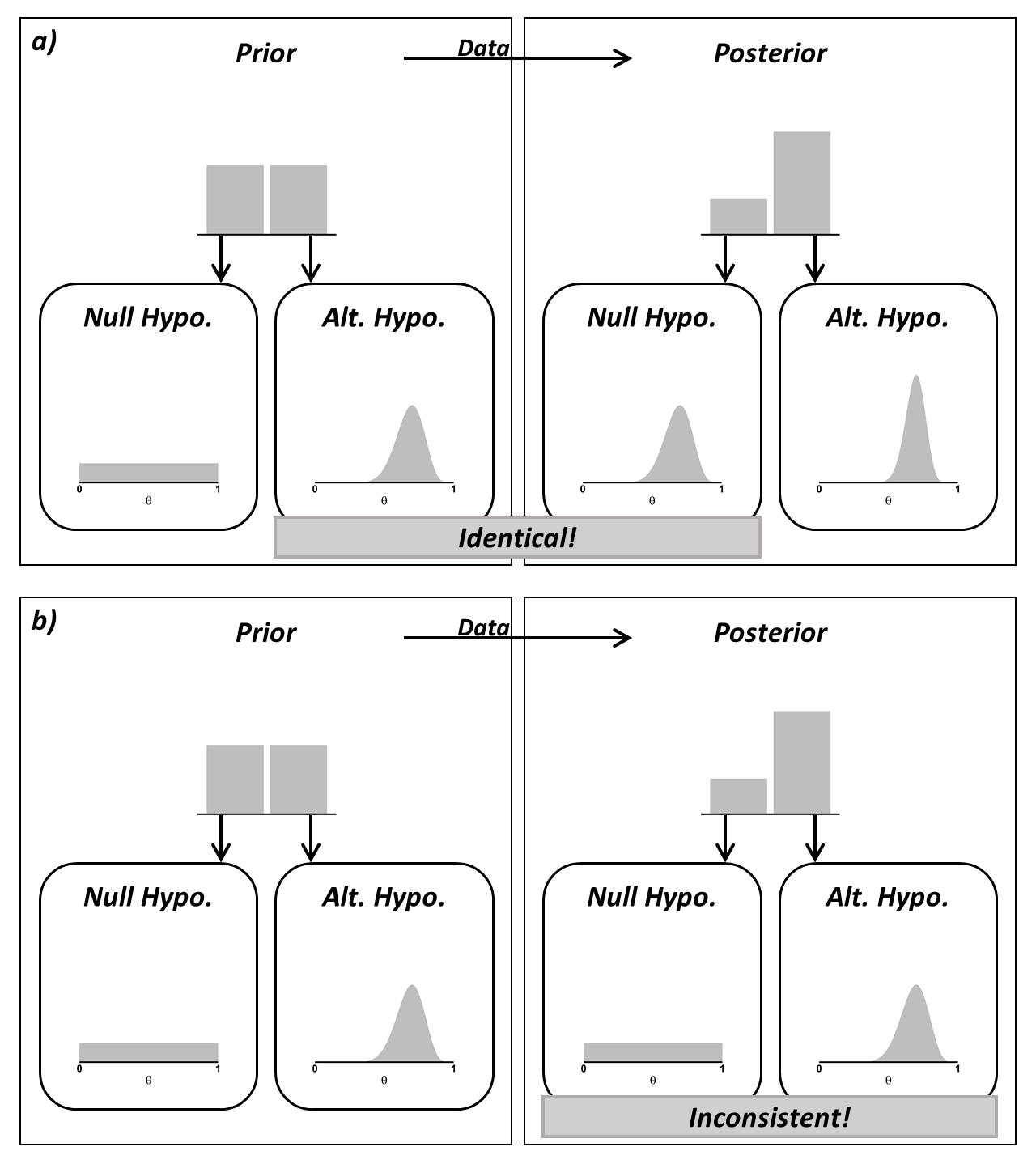}
	\caption{Example. Theoretical positions are represented by parameter distributions with consistent (a) and inconsistent (b) Bayesian updating. Consistent updating yields a situation in which the prior alternative hypothesis and the posterior null hypothesis represent the same theoretical position \citep[This illustration of the hypothesis-based parameter distributions was inspired by][]{Kruschke2018b}.}
	\label{fig:exp_beta}
\end{figure}

An example shall be provided that leads to a paradox with the representation of hypotheses via parameter distributions. Assume the data $x$ is characterized by the binomially distributed random quantity $X \sim Bin(n, \theta)$, with $n=20$ and $\theta \in [0,1]$ (probability of success) being the parameter of interest. Both hypotheses shall have an identical support $\Theta_0 = \Theta_1 = [0,1]$, but different prior (beta) distributions (see Figure~\ref{fig:exp_beta} left, rounded boxes):
\begin{equation}
H_0: \theta \sim Beta(1,1) \quad \text{vs.} \quad H_1: \theta \sim Beta(15,7) \, .
\end{equation}
Further, both hypotheses shall have equal prior probabilities $p(H_0) = p(H_1) = 0.5$ (see Figure~\ref{fig:exp_beta} left).

Now, assume that $s=14$ successes were observed. The resulting Bayes factor is $B\!F = 2.89$ leading to posterior probabilities $p(H_0|s) = 0.257$ and $p(H_1|x) =0.743$, favoring the alternative hypothesis $H_1$ (see Figure~\ref{fig:exp_beta} right).
However, also the within-hypothesis beta distributions get updated by observing $s$ to (see Figure~\ref{fig:exp_beta}a, right, rounded boxes; else updating is inconsistent (see Figure~\ref{fig:exp_beta}b))
\begin{equation}
H_{0|s}: \theta|s \sim Beta(15,7) \quad \text{vs.} \quad H_{1|s}: \theta|s \sim Beta(29,3) \, .
\end{equation}
This example was constructed such that the posterior null hypothesis $H_{0|s}$ has the same distribution as the prior alternative hypothesis $H_1$. If a theoretical position is represented by a parameter distribution, then both of these hypotheses represent the same theoretical positions (Figure~\ref{fig:exp_beta}a). Although the prior alternative hypothesis $H_1$ \textit{gains} credibility by observing $s$, the posterior null hypothesis $H_{0|s}$ has \textit{less} credibility due to observing $s$. Paradoxically, both hypotheses represent the same theoretical position, so it is not clear whether the data agree or disagree with this theoretical position. 

In order to solve this paradox, hypotheses need to be considered as sets of parameters. Both hypotheses hypothesize the same parameter set, representing the same theoretical position. Accordingly, the research question, that will be answered within this analysis, contrasts a theoretical position against itself. Naturally, no data can inform this pointless contrast.

\section{Discussion}

This paper elaborated that hypotheses should be represented by sets of parameters only, not by parameter distributions. If so, updating consistency and a proper evidential interpretation of Bayes factors are given, and if not, the foundational or evidential basis of Bayes factors is severely impaired.
In that, a clear distinction between theoretical positions and knowledge about the phenomenon of interest is mandatory. It is important that the content of theoretical positions should only inform the specification of hypotheses (sets of parameters) and that the available knowledge should only inform the specification of the prior distribution.

\subsection{Empirical Content}

It is argued that by using parameter distributions to represent hypotheses, their \textit{empirical content} can be increased \citep[][p.~1052]{Vanpaemel2012}, supplemented by references to \citet[][]{Popper2002}. However, the elaborations within this paper might cast doubt. According to \citet[][p.~103]{Popper2002}, the empirical content of a statement is ``the class of its potential falsifiers'', such that a higher empirical content is characterized by a larger class of potential falsifiers. Now, one needs a proper concept of ``falsifiability'' in the Bayesian framework, and it appears that defining such a concept is fundamentally difficult, as Popper's elaborations of induction are restricted to the \textit{modus tollens} of deductive tests \citep[][p.~19]{Popper2002} while the Bayesian framework tries to formalize induction by a logic of partial beliefs \citep[cp.\ e.g.][p.~20]{Ly2016}. If, at all, one tries to find such a concept, one might say that a hypothesis is falsified if its probability is zero. Naturally, a zero probability of a hypothesis will not be obtained in a scientific investigation (which assumes non-zero prior probabilities), so -- practically -- one might stop if the probability of a hypothesis is sufficiently small and then decide to treat this hypothesis as falsified. In terms of the limit behavior of Bayes factors, this resembles the case in which the Bayes factor tends towards $\infty$ or $0$ \citep[cp.\ also][p.~105]{Rouder2018}. This refers to the first two cases in equation (\ref{eq:BF_inf}). In the third case, however, evidence will never be conclusive if $n \to \infty$, so it will not be possible to ``falsify'' any of the hypotheses with the given experiment.
In that, the class of potential falsifiers of $H_0$ is $\Theta_1 \setminus \Theta_0$ and the class of potential falsifiers of $H_1$ is $\Theta_0 \setminus \Theta_1$.
Therefore, it is only the supports $\Theta_0$ and $\Theta_1$ that determine the empirical content of the hypotheses, not the exact shape of the prior distributions ($\pi(\theta|H_0)$ and $\pi(\theta|H_1)$). Consequently, beyond their mere supports, prior distributions do not increase the empirical content of hypotheses.

\subsection{Nil-Hypotheses}

Acknowledging that hypotheses are only the supports of the within-hypothesis prior distributions, it appears that many elaborations in the context of Bayes factors \citep[e.g.][]{Goenen2005,Rouder2009,Rouder2018,Rouder2018b,Dienes2019} do still use a sharp null hypothesis, which hypothesizes only one single parameter value \citep[cp.\ also][p.~787]{Tendeiro2019}. In that, these hypotheses are identical to those employed in conventional null hypothesis significance testing (NHST), such that its heavy critique about the uselessness of these hypotheses \citep[see e.g.][]{Berkson1938,Cohen1994,Kirk1996,Gigerenzer2004} does apply to these Bayes factors as well. The inclusion of the parameter distribution into the statistical analysis does not tackle these issues (about the uselessness of the employed hypotheses). To do so, hypotheses need to be specified as sets of parameters that correspond to the theoretical positions that are of interest within the research question. Then, these hypotheses are typically not single-valued anymore. In this regard, the methodological development of Bayes factors with reasonably specified interval-valued hypotheses needs to be addressed more intensively. Although few elaborations exist \citep[cp.][]{Morey2011,Hoijtink2019, Heck2020}, this development is treated as rather ancillary within the Bayes factor literature. Alternative hypothesis-based methods \citep[see e.g.][]{Lakens2017, Lakens2018, Kruschke2015, Kruschke2018} already started to primarily address this necessity of allowing reasonably specified interval-valued hypotheses, and Bayes factors need to go along with them.

\subsection{Knowledge vs. Theoretical Positions}

The central message of this paper is that knowledge and theoretical positions about the phenomenon of interest need to be distinguished. Former inform the specification of the prior distribution, latter inform the specification of the hypotheses. In that sense, both of these mathematical objects (prior distribution, hypotheses) or real world concepts (knowledge, theoretical positions) are independent of each other. This can also be seen, as it is possible to specify a prior distribution without having hypotheses (as in a non-hypothesis-based Bayesian analysis) or as it is possible to specify hypotheses without having a prior distribution (as in non-Bayesian hypothesis-based analyses). Yet, it is possible to depict the prior distribution in dependence of the hypotheses via the within-hypothesis prior distributions (equations (\ref{eq:pi_h0}) and (\ref{eq:pi_h1})) and the prior probabilities of the hypotheses (equation (\ref{eq:prior_probs})). Strikingly, after combing these components to the overall prior distribution (equation (\ref{eq:prior})), its dependence on the hypotheses is gone! Naturally, what is known about the phenomenon of interest does primarily not depend on which hypothetical conjectures might be possible about it. This has serious implications about how to specify the essential quantities in a hypothesis-based Bayesian analysis: It is recommended to specify the overall prior distribution (as density $\pi(\theta)$) and the hypotheses (via $\Theta_0$ and $\Theta_1$) independently. If, in contrast, the within-hypothesis prior distributions shall be specified (via $\pi(\theta|H_0)$ and $\pi(\theta|H_1)$), the applied scientist needs to make sure that by combining them with the prior probabilities of the hypotheses to the overall prior distribution (equation (\ref{eq:prior})) its dependence on the hypotheses is gone. This seems quite remarkable.

\subsection{Outlook}

Looking at the history of Bayesian statistics, it appears that prior distributions have always had a bad reputation. In this context, it seems that the idea of using prior distributions to formalize theoretical positions was motivated by the intention of correcting this bad reputation of prior distributions. For example, \citet[both quotes on p.~1048]{Vanpaemel2012} stated that they ``do not agree that priors are an unwanted aspect of the Bayesian framework'' and that they ``believe that it is wrong to malign priors as a necessary evil''. It can only be agreed on! Parameter distributions are a vital part of Bayesian statistics and must not be condemned!
This elaboration clarified the distinction between knowledge (parameter distribution) and theoretical positions (hypotheses), and, therefore, tried to contribute to a correct employment of parameter distributions in the Bayesian framework.

\bibliography{Preprint_BayesianHypotheses}

\end{document}